\documentclass[aps,preprint]{revtex4}
\usepackage{graphicx,amssymb}
\begin{document}
\setcounter{page}{1}

\title{ 
 Increases of a Diamagnetic Property by Flux-Pinning in Volume Defect-Dominating  Superconductors}

\author{H. B. \surname{Lee}}
\email{superpig@pusan.ac.kr}
\thanks{Fax: +82-51-513-7664}
\author{G. C. \surname{Kim}}
\author{Y. C. \surname{Kim}}

\affiliation{Department of Physics, Pusan National University, Busan 46241, Korea}
\author{R. K. \surname{Ko}}
\author{D. Y. \surname{Jeong}}

\affiliation{Korea Electrotechnology Research Institute, Changwon 51543, Korea}

\begin{abstract}
Whereas there are two critical fields that are H$_{c1}$ and H$_{c2}$ in the ideal type II superconductor,  there is another critical field H$_{c1}'$ defined as the field showing the maximum diamagnetic property  
in the real type II superconductor. 
 We would present that H$_{c1}'$ is able to be proved 
  theoretically and experimentally. 
  We have derived an equation based on flux-pinning effect of  volume defects. 
MgB$_2$ bulks which were synthesized by Mg and B are similar to this model. 
 The number of quantum fluxes pinned at a defect of radius r, a pinning penetration depth,  magnetic flux penetration method, and a magnetization at H$_{c1}'$ in the static state are suggested  through the equation of the model. It was speculated  that  pinned fluxes at a volume defect in the superconductor have to be pick-out depinned from the defect and move an inside of the superconductor when  pick-out forces of pinned fluxes  is larger than  pinning force of the defect (F$_{pickout}$ $>$ F$_{pinning}$) or when the shortest distance between pinned fluxes at a volume defect is the same as that of H$_{c2}$. 
In reality, $\Delta$G$_{dynamic}$ which is sum of fluxes movement energy and fluxes vibration energy is involved in movement of pinned fluxes. When  volume defects are small and many, the number of pinned fluxes at a volume defect calculated by experimental results was closer to  that of ideally calculated ones because of a small $\Delta$G$_{dynamic}$.  However, when volume defects are large and a few, the number of pinned fluxes at a volume defect calculated by experimental results were much fewer than  that of ideally calculated ones because of a large $\Delta$G$_{dynamic}$.

\end{abstract}



\maketitle

\section{Introduction}  
One of phenomena which have not been clearly explained  in superconductors is that  diamagnetic property of superconductors increases after H$_{c1}$.  Field dependences of magnetization of superconductors (M-H curves) show that most of superconductors increase their diamagnetic property by rounded form after H$_{c1}$. 
 This is against the definition  that superconductors decrease its diamagnetic property after H$_{c1}$. 
Although the increase of the diamagnetic property  after H$_{c1}$ is obvious in almost all of superconductors,  superconductor researchers accept it implicitly  because  most of superconductors showed the general behavior superficially. 



Generally, it has been understood that there is H$_{c1}$ and H$_{c2}$ in the type II superconductor \cite{Tinkham3, Radzyner, Sugui, Lee2}. However, this definition is for the ideal type II superconductor. It is certain that there is H$_{c1}'$ which is defined as the field showing maximum diamagnetic property in the real type II superconductor because flux-pinning effects of defects  exist in superconductor. 
 Although many researchers have studied for flux-pinning effects,  
 their results  depend only on experimental techniques owing to the absence of proper theory \cite{Shabbir, Malmivirta, WeiZhou}. 
 The purpose of this paper is to present a possibility that H$_{c1}'$ would be an another critical field in superconductors and relative detail mechanisms. 

 In fact, it is not easy to indicate  H$_{c1}$ in  M-H curves whereas it is simple to indicate  H$_{c1}'$, 
which is just the applied magnetic field of the maximum diamagnetic property in a superconductor.  According to the definition of type II superconductor, M-H curve must be a straight line to H$_{c1}$ and  diamagnetic property of the superconductor must decrease owing to flux penetration if applied field increases over H$_{c1}$. 
 However, all of real type II superconductor do not reduce their diamagnetic property although applied field  passes over H$_{c1}$. 
They show much better diamagnetic property at H$_{c1}'$ than H$_{c1}$. It is certain that diamagnetic property increase if penetrated fluxes are pinned at volume defects and do not move an inside of the superconductor. The behavior means that volume defects in a superconductor are another barrier to prevent the fluxes from penetrating into the superconductor.
 
Many researchers have stated that H$_{c1}$ of MgB$_2$ is 250 - 480 Oe \cite{Askerzade, Buzea}. However, the maximum diamagnetic property appears at much higher fields. In our experiments, diamagnetic property in pure MgB$_2$ increases over 5000 Oe. The behavior was  shown in other MgB$_2$ superconductors,  melt-textured growth (MTG) specimens  and single crystal superconductor (SCC), which can be called as a volume defect-dominating superconductor (VDS) \cite{SLLi, Wei, Yadav}. The phenomenon appears not only in VDSs, but also in planar defect-dominating superconductors (PDS) such as  high temperature superconductors (HTSC) which were made by solid state reaction method \cite{Anupam}. 

A study for an increase of diamagnetic property after H$_{c1}$ is that the one of  Bean-Livingston (B-L)  is  unique \cite{Livingston}. Most of superconducting single crystal (SSC) researchers have explained the increase of diamagnetic property after H$_{c1}$ based on the study \cite{Westerholt, Konczykowski}. 
However, the phenomenon is not limited to SSC, but rather higher in bulks as mentioned. Although B-L condition (a clean surface) is unsatisfied in bulk superconductors, there is an increase of diamagnetic property than that of H$_{c1}$. Thus, the effect of B-L and flux-pinning phenomena  are concurrently or competitively occurring in SSCs for the increase of diamagnetic property over H$_{c1}$ because SSCs are not completely free from defects.

On the other hand, there is demagnetization factor in superconductors, which means that  diamagnetic property  is changed  depending on its geometry \cite{Zeldov, Yamasaki,  Brandt}. We did not consider demagnetization factor theoretically and experimentally.  The main concern is the path that diamagnetic property increase was different according to the number of volume defects in MgB$_2$ although geometry of samples are  same. 
Therefore, we cannot help thinking that the behavior was caused by flux-pinning effects of volume defects and would introduce relative studies.
\section{Experiment}
  (Fe, Ti) particle-doped MgB$_{2}$ specimens were synthesized using the non-special atmosphere synthesis (NAS) method \cite{Lee4}. 
 The starting materials were Mg (99.9\% powder) and B (96.6\% amorphous powder) and (Fe, Ti) particles. Mixed Mg and B stoichiometry, and  (Fe, Ti) particles were added by weight. They were finely ground and pressed into 10 mm diameter pellets. (Fe, Ti) particles were ball-milled for several days, and average radius of (Fe, Ti) particles was approximately 0.163 $\mu$m. 
   On the other hand, an 8 m-long stainless- steel (304) tube was cut into 10 cm pieces. One side of the 10 cm-long tube was forged and welded. The pellets and excess Mg were placed in the stainless-steel tube. The pellets were annealed at 300 $^o$C  for 1 hour to make them hard before inserting them into the stainless-steel tube. The other side of the stainless-steel tube was also forged. High-purity Ar gas was put into the stainless-steel tube, and which was then welded. Specimens had been synthesized at 920 $^o$C  for 1 hour and cooled in air. Field dependences of magnetization were measured using a MPMS-7 (Quantum Design). During the measurement, sweeping rates of all specimens were made equal for the same flux-penetrating condition.

\section{results}

\subsection{Experimental confirmation of diamagnetic property increases after H$_{c1}$}

Figure \ref{fig0} shows diamagnetic behaviors of two different (Fe, Ti) particle-doped MgB$_2$ specimens with various temperature, which are fabricated by the same condition except for dopant level of (Fe, Ti) particles. Figure \ref{fig0} (a) is  magnetization behaviors of 5 wt.$\%$ (Fe, Ti) particle-doped MgB$_2$ (5 wt.\% specimen) and shows different H$_{c1}$ with various temperature.  The deviation point from the linear line is the H$_{c1}$. Calculated roughly from the figure,  H$_{c1}$ of 5 wt.\% specimen at 0 K is almost same as that at 5 K, which is 600 Oe. Figure \ref{fig0} (b) is diamagnetic behaviors of 25 wt.$\%$ (Fe, Ti) particle-doped MgB$_2$ (25 wt.\% specimen) with various temperatures. Their behaviors are similar with that of 5 wt.\% specimen in  the viewpoint that diamagnetic property increases after  H$_{c1}$.
 
  On the other hand, H$_{c1}$ of 25 wt.$\%$ specimen at 0 K is approximately 500 Oe which is rather lower than the one of 5 wt.$\%$ specimen. The decrease of H$_{c1}$ can be interpreted as a disturbance of the circulating current by over-doping of (Fe, Ti) particle.  M-H curves of 5 wt.$\%$ specimen and  25 wt.$\%$ specimen to 3 kOe are shown in  Fig. \ref{fig0} (c) and (d), respectively. After reaching H$_{c1}'$ in both figures, it is observed that  $\Delta$H=$\Delta$B regions appears, which is the region that an increased applied magnetic field is the same as the increased magnetic induction \cite{Lee}. In  a volume defect-dominating
superconductor, we represented that diamagnetic property does not decrease  after reaching the maximum diamagnetic property, 
but decreases after forming $\Delta$H=$\Delta$B region when volume defects are many enough. Figure \ref{fig0} (c) and (d) support the representation.


\subsection{A theoretical view of H$_{c1}'$}
 When magnetic field is applied on a superconductor, which has m$'^3$ volume defects of radius r as shown in Fig. \ref{fig1} (a), quantum fluxes are pinned on the defects after H$_{c1}$ as shown Fig. \ref{fig1} (b), (c), and (d). 
When a magnetic field $H$ is applied  over H$_{c1}$, 
the number of quantum fluxes (n$^2$) pinned at a defect of radius r  in the static state is 
\begin{eqnarray}
n^4 - \frac{2H^2}{\alpha} r^3   = 0
\end{eqnarray} 
where  $\alpha$ is $\frac{aLH_{c2}\Phi_o}{\sqrt{P}c}$ and $r$ is radius of a volume defect. $\Phi_o$ is flux quantum which is 2.07$\times10^{-7}$ G$\cdot$cm$^2$, 
$aL$ is an average length of  quantum fluxes which are pinned and bent between volume defects (a is an average bent constant which is $1<a<1.2$ and $L$ is the distance between defects in vertically packed state), 
and $P$ is the filling rate which is $\pi/4$ when flux quanta are pinned at the volume defect in the form of square \cite{Lee}.  In reality, the number of flux quanta  pinned at a spherical volume defect of radius $r$ is $n$$^2$.

The equation shows how many quantum fluxes are pinned at a volume defect of which radius is $r$ in a superconductor when a magnetic field is applied in diamagnetic increase region.  
We postulate a theory in the assumption that quantum fluxes pinned at a volume defect in the superconductor are pick-out depinned from the defect and move an inside of the superconductor due to two following reasons.

 The one is  when pick-out forces of pinned fluxes are larger than the pinning force of the defect (F$_{pickout}$ $>$ F$_{pinning}$).  And the other is when the shortest distance between quantum fluxes pinned at a volume defect are the same as that of H$_{c2}$. The justification of the latter came from the fact that there is no flux-pinning effect if the neighborhoods of a volume defect are not a superconducting state anymore.

Deviating from H$_{c1}$ linear line means that magnetic fluxes penetrated into an inside of superconductor. And an increase of diamagnetic property after H$_{c1}$ means that fluxes having penetrated are pinned on volume defects and block the external magnetic field. Diamagnetic property of a superconductor would continue increasing as H is increasing if there are a lot of volume defect.  It is because penetrated fluxes still remain near the surface of superconductor due to a lot of volume defects. However, we noticed that there is a limit of pinned fluxes at a volume defect, which is that fluxes pinned at a volume defect definitely pick-out depinned and move into an inside of the superconductor regardless of  $\Delta$G$_{defect}$ if the short distance (d$'$) between quantum fluxes pinned at a volume defect is the same as that of  H$_{c2}$. $\Delta$G$_{defect}$ is defined as free energy density difference between a volume defect and superconductor. 

 H$_{c1}'$ is the field that pinned fluxes at a volume defect have to be pick-out depinned and move regardless of  $\Delta$G$_{defect}$. Experimentally,  H$_{c1}'$ on a M-H curve appears as a vertex or as a broad region which is $\Delta$H=$\Delta$B region.   H$_{c1}'$ is the vertex of M-H curve in most of superconductors,, but some of the volume defect-dominating superconductors showed a broad region of H$_{c1}'$ in the M-H curve if volume defects in the superconductor is appropriate  \cite{Lee, Hsu, Muller}.
 In our experiment, it is clearly observable  in Fig. \ref{fig0} (c) and (d). 
After reaching H$_{c1}'$ in both figures, it is observed that $\Delta$H=$\Delta$B region which means broad H$_{c1}'$ appeared \cite{Lee}. 
 
Penetrating behaviors of magnetic fluxes to H$_{c1}'$ in a superconductor 
are explained in Fig. \ref{fig2-6}. They are calculated by Eq. (1) in the condition that the number of  volume defects of which radius is 163 nm  are  4000$^3$  in 1 cm$^3$.  H$_{c1}$ and H$_{c2}$  are assumed as 400 Oe and 50 Tesla (T) at 5 K, respectively \cite{Lee}. When H is raised to 450 Oe, calculated numbers of quantum fluxes pinned at a volume defect are 28$^2$ and penetrated fluxes are pinned to third defects from the surface. According to increases of applied field (H), penetrated quantum 
fluxes are going to not only  move another pinning site,
but also increase the number of pinned 
fluxes on defects. Each volume defect which has pinned fluxes reaches its pinning limit (n$^2$=45$^2$) when H is 1435 Oe which is H$_{c1}'$, as shown in Fig. \ref{fig2-6} (e). More H induces  $\Delta$H=$\Delta$B region  because volume defects already reached their pinning limit (Fig. \ref{fig2-6} (f)).

\subsection{The derivation of the equation}
If a magnetic field $H_{c1}$ is applied to the superconductor,  a free energy (FE) density  of a superconductor will be raised to  $\Delta G_s$ = $H_{c1}^2/8\pi$.
 On the other hand,  FE density of a defect in the superconductor will be $\Delta G_n$ = 0 at $H_{c1}$ because there is no magnetic fluxes in the defect and it is not superconductor. 
  Magnetic fluxes begin to penetrate into the specimen beyond the $H_{c1}$ through the surface of the superconductor  in 
the state of flux quantum. 
 Quantum fluxes which have penetrated into the superconductor are willing to move into an inside of specimen by a concentration gradient of quantum fluxes and repulsive force between them, and will be pinned at the defect where FE density is lower than that of the superconductor. In this situation, fluxes had penetrated into the superconductor are pinned at the defect around the surface of the superconductor if volume defects are many enough. 
 If  an applied field is raised to $H$ which is parallel to the specimen surface,  FE density difference between superconductor and normal state when penetrated  magnetic fluxes after $H_{c1}$ are pinned at volume defects is  
 {\setlength\arraycolsep{2pt}
\begin{eqnarray}
\Delta G_n - \Delta G_s = - \frac{(H-B)^2}{8\pi} 
\end{eqnarray}
where B is magnetic induction which are pinned at volume defects after $H_{c1}$
 
Therefore, FE density of a spherical  defect is 
{\setlength\arraycolsep{2pt}
\begin{eqnarray}
\Delta G_{defect} = -  \frac{(H-B)^2}{8\pi}\times \frac{4}{3}\pi r^3+ \Delta\epsilon_{n} \times 2r 
\end{eqnarray}
 where $H$  is an applied field, $r$ is a radius of defect and 
$\Delta\epsilon_{n}$ is  FE density increase of the defect by pinned fluxes.  Generally, B can be ignored in diamagnetic increase region because they are too small.  Since the fluxes in the superconductor exist in the state of quantum, they have vortexes of  eddy currents. Hence $\Delta\epsilon_{n}$ is  $n^2\Phi_o^2/8\pi$, 
where  $n$ is the number of flux quanta pinned at the defect and  flux quantum $\Phi_o $ is 2.07$\times10^{-7}$G$\cdot$cm$^2$.  Multiplying $2r$ to $\Delta\epsilon_{n}$ in Eq. (8) means that  FE density of a defect increases as much as $2r$ in the pinned quantum fluxes at a volume defect.  
Looking at a depth of FE density of a volume defect, it is entirely dependent on a volume of the defect. 

 On the other hand, total forces acting on fluxes pinned at a volume defect are softened pinning forces by  quantum fluxes stacked on the defect and tension forces acting between quantum fluxes pinned at the defect. The tension force of the forefront flux quantum  in two dimensions (2 D), as shown in Fig. \ref{fig1} (c), is the number of quantum fluxes which are pinned at the volume defect times a repulsive force acting between quantum fluxes. 
 
The repulsive force per unit length (cm) between quantum fluxes, which is caused by vortexes, is
{\setlength\arraycolsep{2pt}
\begin{eqnarray}
f = J_s\times\frac{\Phi_o}{c}    
\end{eqnarray} 
where $J_s$ is the total supercurrent  density due to vortices \cite {Tinkham3}. 
Assuming that $n$ quantum fluxes are pinned between two defects of which distance is $L$, the tension force on the forefront of the quantum fluxes is
{\setlength\arraycolsep{2pt}
\begin{eqnarray}
f_{tension} = 
aLnJ_s\times\frac{\Phi_o}{c}    
\end{eqnarray} 
where $n$ is the number of quantum fluxes which are pinned horizontally at a volume defect, $aL$ is a average length of  quantum fluxes which are pinned and bent between defects (a is average bent constant, $1<a<1.2$)  and $J_s$ can be expanded  to total 
 supercurrent density. 
  These behaviors are explained in Fig. \ref{fig1} (b) and (c). 

There are several reports for flux-pinning mechanism in superconductor  \cite{Doria, Latimer, Willa, Chen, Golosovsky}, but there has been no report for flux-pinning of volume defects in superconductors.  We derived equations in the assumption that movements of pinned fluxes at volume defects in superconductor would obey Hook's law. 
A tension force of quantum fluxes which are pinned between two defects  can be described as a behavior of Hook$'$s law \cite{ Chen, Golosovsky}
{\setlength\arraycolsep{2pt}
\begin{eqnarray}
f_{tension} = - kx   
\end{eqnarray}
 where $k$ is a quantum flux tension constant and $x$ is the displacement 
 from 
the center 
 of two defects which pinned  quantum fluxes to the forefront quantum flux as shown  
 in Fig. \ref{fig1} (c).  
The displacement $x$ can be expressed as
{\setlength\arraycolsep{2pt}
\begin{eqnarray}
x = n\bar d, \bar d= \frac{\sum_{n=1}^n d_n}{n}
\end{eqnarray} 
where d$_n$ is  the maximum distance among distances between pinned fluxes as defined in the Fig. \ref{fig1} (b). Therefore, the quantum flux tension constant $k$ is 
{\setlength\arraycolsep{2pt}
\begin{eqnarray}
k = \frac{aLJ_s\Phi_o}{\bar{d}c}  
\end{eqnarray} 
On the other hand, there might be quantum fluxes which are not pinned at any volume defect when the distance between defects is wide enough. These unpinned fluxes would be present in a compressed state because there are pinned fluxes before them. Thus, they cause the pushing force (F$_ {pushing}$) to push pinned fluxes forward. 

 Quantum fluxes pinned at a volume defect begin to move by being pick-out depinned  if the pinning force and the pushing plus tension forces are same. Thus,
{\setlength\arraycolsep{2pt}
\begin{eqnarray} 
\Delta F_{pinning}=\frac{\partial G}{\partial r} = - \frac{(H-B)^2}{8\pi}\times 4\pi r^2 + \frac{2n^2\Phi_o^2}{8\pi} 
\end{eqnarray}
Assuming that Q times fluxes of  the number of fluxes pinned at a volume defect exist between defects in the compressed state,
\begin{eqnarray}
\Delta F_{pickout} 
   =- Qn_vLnJ_{s1}\frac{\Phi_o}{c} - n_v\sum_{n=1}^n kx_n 
=- Q\frac{LJ_{s1}\Phi_o}{c}n^2 - \frac{aLJ_s\Phi_o}{2c}n^3
\end{eqnarray} 
where $J_{s1}$ is the current density of circulating flux quantum in a compressed state.
It was assumed that the distance between quantum fluxes pinned at a volume defect to the vertical direction is the same as that of horizontal direction  
and 
 $n_v$ (the number of quantum fluxes vertically pinned at the defect)
is same as $n$, thus fluxes pinned at the defect are vertically subjected to the same tension. They will be pick-out depinned and move into an inside of the superconductor if $\Delta$F$_{pickout}$ is larger than $\Delta$F$_{pinning}$.
 \begin{eqnarray} 
- \frac{H^2}{8\pi}\times 4\pi r^2 + \frac{2n^2\Phi_o^2}{8\pi}
 = -Q \frac{LJ_{s1}\Phi_o}{c}n^2 -\frac{aLJ_s\Phi_o}{2c}n^3
\end{eqnarray}

On the other hand, the critical current density (J$_c$) between two flux quanta can be calculated as Silsbee criterion \cite{Tinkham4}.
  \begin{eqnarray}
 J_c =\frac{c}{4\pi}\frac{N\Phi_o }{d'/2}\leq \mu_0\frac{2H_{c2}}{d'}
 \end{eqnarray} 
 where $N$ is the number of flux quantum in the superconductor and $d'$ is the minimum distance between quantum fluxes pinned at a defect, which is shown in Fig. \ref{fig1} (b).   $J_c$ of eddy current  is the maximum when $d'^2$ is $2\pi\xi^2$.
  
 Since the repulsive force between flux quanta are inversely proportional to $d'$, 
  J$_c$ of eddy current circulating the flux quantum are also inversely proportional to $d'/2$ by Eq. (12). 
 Therefore, J$_{s, d'}$ is
\begin{eqnarray}
 J_{s, d'} =\mu_0\frac{H_{c2}}{d'/2} = \mu_0\frac{nH_{c2}}{\sqrt{P} r}
 \end{eqnarray} where 
 $r$ is the radius of a volume defect, $n^2$ is the number of quantum flux pinned at a volume defect and $P$ is filling rate which is $\pi$/4 when pinned fluxes have square form. 
 During the calculation, $d'/2$ is derived from the following equation when flux quanta are pinned at a defect in the form of square \cite{Lee}.
\begin{eqnarray}
n^2 = \frac{\pi r^2}{\pi (\frac{d'}{2})^2}\times P 
 \end{eqnarray}  
 The average value of $J_s$ between $d$ and $d'$ have to be inserted into Eq. (11), which is almost J$_{s, d'}$/2. 
 Assuming that J$_{s1}$ is $\frac{1}{3}$J$_{s, d'}$, the equation is 
 \begin{eqnarray}
\frac{aLH_{c2}\Phi_o}{4c\sqrt{P}}n^4 + Q \frac{L H_{c2}\Phi_o}{3c\sqrt{P}}n^3 - \frac{H^2}{8\pi}\times 4\pi r^3  = 0
\end{eqnarray} The term $\frac{2n^2\Phi_o^2}{8\pi}$ was ignored owing to $\Phi_o^2$.
 Putting $\alpha$=$\frac{aLH_{c2}\Phi_o}{\sqrt{P}c}$
 \begin{eqnarray}
n^4 + \frac{4Q}{3a} n^3 - \frac{2 H^2}{\alpha} r^3   = 0
\end{eqnarray} 
When Q is zero,
\begin{eqnarray}
n^4 - \frac{2H^2}{\alpha} r^3   = 0
\end{eqnarray}

\subsection{Pinning force of  a volume defect,  pinning penetration depth, and  diamagnetic property at H$_{c1}'$}

Pinning force of a defect (f$_r$) at H$_{c1}'$ is calculated from Eq. (15) in Method  by inserting the number of flux quanta (n) of Eq. (1) when Q is zero and shown in the Fig. \ref{fig2-7} (a) and (b) under the condition that H$_{c1}$ is 600 Oe. Figure \ref{fig2-7} (a) is a pinning force distribution at H$_{c1}'$ along the number of volume defects in the superconductor when the radius of defect is 163 nm. Despite the same size of defects, pinning force of a defect shows a significant difference according to the number of volume defects in a superconductor. It is observed that the pinning force of a volume defect  greatly increased as the number of volume defects decrease. Figure \ref{fig2-7} (b) is a pinning force distribution at H$_{c1}'$ along the number of volume defects in the superconductor when vol.\% of the defects is fixed as 1.56.  This is the case that the volume defects keep being divided to a smaller size.  As the number of volume defects increases (their size decreases), the pinning force of a volume defect decreases dramatically. 

 Considering  pinning penetration depth, which is defined as the depth of pinned fluxes on the defects in the superconductor at H$_{c1}'$, the force which push the quantum fluxes into an inside of the superconductor by applied field must be equal to sum of the bulk pinning force by Meissner effect and pinning forces of volume defects  from the surface of the superconductor. It is assumed that quantum fluxes are penetrating into the superconductor along x-axis  (quantum fluxes are laid on the y-axis) and an yz plane has $m'^2$ defects as shown Fig. 2 (a) and (b).
 \begin{eqnarray} 
\frac{\partial({G_{H_{c1}'} - G_{H_{c1}})}}{\partial m} =m m'^2f_r 
\end{eqnarray} where $f_r$ is the pinning force of a defect  of  which radius is $r$ and $m$ is the number of volume defects which have pinned fluxes to the flux movement direction. Thus, 
\begin{eqnarray} 
 \frac {H_{c1}'^2 - H_{c1}^2}{8\pi}=\frac{m^2}{2}m'^2f_r  
 \end{eqnarray}

 Therefore,  the pinning penetration depth of a specimen at H$_{c1}'$ is $x=mL'$, where $L'$ is the distance between defects in regular state. Calculated pinning penetration depths for the various states are shown in Fig. \ref{fig2-7} (c) and (d) under the condition that H$_{c1}$ is 600 Oe. Figure \ref{fig2-7} (c) is the calculated pinning penetration depths at H$_{c1}'$ along the number of volume defect when the radius of defect is 163 nm. It is observed that pinning penetration depth increases as the number of volume defects decreases.  A calculated pinning penetration depths at H$_{c1}'$ along the number of volume defect when the vol.\% of the defects is set as 1.56 are shown in Fig. \ref{fig2-7} (d).  At the same vol.\% of the defects,  the fluxes do penetrate deeply at H$_{c1}'$ as the radius of defect increases.

 Magnetic induction in a superconductor which have m$'^3$ volume defects is
 \begin{eqnarray}
B =n^2m_{cps}m\Phi_o
 \end{eqnarray} 
where $n^2$, $m_{cps}$, and $m$ are the number of quantum fluxes pinned at a defect,  the number of defect which is vertically closed packed state, and the number of volume defects which had pinned fluxes from the surface of the superconductor along x-axis as shown in Fig. \ref{fig1} (a), respectively. The  $m_{cps}$ is the minimum number of volume defects if  fluxes penetrated  into the superconductor are completely pinned ($m_{cps}$$\times 2r =1$) \cite{Lee}. Thus, the total numbers of flux quanta pinned on volume defects of a plane perpendicular to the flux moving direction are $n^2m_{cps}$.

 The diamagnetic property M at H$_{c1}'$ is
  \begin{eqnarray}
  M = \frac{B-H}{4\pi} = \frac{ n^2m_{cps}m\Phi_o - H}{4\pi}
 \end{eqnarray} where H is H$_{c1}'$.
  Calculated diamagnetic property at H$_{c1}'$ in the static state 
are shown in Fig. \ref{fig7}.  
 
\section{Discussion}
The number of flux quanta pinned at a volume defect has been calculated in a static state. However, two terms are added in the dynamic state.  Magnetic fluxes in volume defect-dominating superconductor are approaching a volume defect with their velocity because they are pick-out depinned together when they are depinned, thus there would be a kinetic energy. And there would be a vibration energy because fluxes pinned on  volume defects are continuously vibrating by movements of other part of quantum fluxes. Magnetic quantum fluxes can be classified as pinned parts at the volume defect and unpinned parts in superconductors, and the latter are continuously affected by movements of other parts of  quantum fluxes. For examples, a flux quantum in 1 cm$^3$ of pure MgB$_2$ are pinned on approximately 1000  volume defects simultaneously.  Thus, the former continue to vibrate because they are interconnected. 
Therefore,
\begin{eqnarray}
\Delta G_{pinning} =\Delta G_{defect} - (\Delta K.E. + \Delta G_{vibration}) = \Delta G_{defect} - \Delta G_{dynamic}
 \end{eqnarray} 

As a distance between  volume defects increases,  $\Delta$$K.E.$ increases because  a large velocity of the quantum fluxes is induced when they are pick-out depinned from the defect. And $\Delta$G$_{vibration}$ increases as the number of flux quanta pinned at a volume defect increase. 
Thus, as the radius of volume defect increases, 
flux quanta would be pick-out depinned from the volume defect at a much fewer flux quanta than that of the static state. We believe in that  $\Delta G_{dynamic}$ would be related with non-equilibrium movement of fluxes (e.g. flux jump). 

On the other hand, as the radius of volume defects decreases and the number of them increases,  $\Delta$G$_{dynamic}$ decreases.  Flux quanta penetrated into the superconductor from the outside, are pinned on the volume defects around the surface, and move into an inside of superconductor before the number of them becomes larger as shown Fig. \ref{fig2-6} and are pinned again on next volume defects. Thus, the diamagnetic property at H$_{c1}'$ is not greatly increased because a pinning limit of the volume defects is low as shown Fig. \ref{fig7} (e) and (f). And the degree of bending of the fluxes is also low because a distance between volume defects is not wide and their pinning limit is low. Therefore, $\Delta$G$_{dynamic}$ is small.

Another problem that breaks an increase of the diamagnetic property in  superconductors is  flux jump. What the number of fluxes pinned  at a volume defect increases means that the number of fluxes moving together also increase because they are moving together when they are pick-out depinned from the defect, which results in generating a lot of heat, and it means a degradation of the superconductor. Therefore, it is believed that flux jumps are dependent on $\Delta$G$_{dynamic}$. 

The maximum diamagnetic property of the pure  MgB$_2$ at 5 K is  approximately  -100 emu/cm$^3$, which is shown in Fig. \ref{fig8} (a). 
The maximum diamagnetic property of pure MgB$_2$ would increase more at 5 K if there were no flux jump. 
As shown in Fig. \ref{fig8} (b),  it is observed that  a higher diamagnetic property  appears at 10 K, which is close  to -150 emu/cm$^3$ when a magnetic field in the opposite direction is applied.  Generally, as temperature of a superconductor decreases,  diamagnetic property increases.  If there were no flux jump at 5 K, it is clear that the diamagnetic property at 5 K would be greater than 10 K. 

We used 96.6 wt.$\%$ purity of the boron in synthesizing pure MgB$_2$, and the impurity of 3.4 wt.$\%$ are  equal to approximately 1.3 vol.$\%$ in MgB$_2$. It produced volume defects of which radius is 1 $\mu$m on average, as shown in Fig. \ref{fig8} (c). It caused flux jump in pure MgB$_2$ specimen by large  $\Delta$G$_{dynamic}$.  In reality,  diamagnetic property would continue to increase over 5000 Oe although H$_{c1}$ of the specimen is approximately 400 Oe. 

A similar calculation for max-diamagnetic property of a superconductor in the static state, which contains volume defects of which radius is 1 $\mu$m and its vol.$\%$ is  1.56, are shown in Fig. \ref{fig7} (c) and (d), which is -456.6 emu/cm$^3$ at 5 K (H$_{c1}'$ is 6370 Oe). 
As the number of volume defects which had pinned fluxes increases, the number of flux quanta pinned at a volume defect increases up to 276$^2$  which is its flux-pinning limit when H$_{c2}$ of the superconductor is assumed to be 50 T. 
However, this phenomenon does not happen because  influences of $\Delta$G$_{dynamic}$ and flux jump are severe in the pure MgB$_2$. In our calculation by Eq. (5), maximum 135$^2$ flux quanta would be pinned at  1$\mu$m volume defects if the max-diamagnetic property have occurred at 5000 Oe in pure MgB$_2$.  And it is observed that they cause a flux jump before reaching H$_{c1}'$ as shown in Fig. \ref{fig8} (a). 

 On the other hand, when (Fe, Ti) particles are doped on MgB$_2$, pinned  fluxes on volume defects have much less opportunity to move a large distances  because (Fe, Ti) particles are pinning fluxes together with 1 $\mu$m radius defects, as shown in Fig. \ref{fig1} (d). We used (Fe, Ti) particles of which  radius is 163 nm on average as the dopant which is  shown in Fig. \ref{fig8} (d).  5 wt.\% (Fe, Ti) particles in MgB$_2$   corresponds to approximately 2.0 vol.\% in MgB$_2$ and means that MgB$_2$ has approximately 8000$^3$ volume defects in 1 cm$^3$. If there were no  defects of which radius is 1 $\mu$m in the superconductor, calculated max-diamagnetic property cannot exceed -52 emu/cm$^3$, as shown in Fig. \ref{fig7} (a) and (b). However, the max-diamagnetic property of the 5 wt.\% (Fe, Ti) particle-doped MgB$_2$ specimen, which  exceed -150 emu/cm$^3$ as shown in Fig. \ref{fig0} (a), is determined as the effect of volume defects of which radius 1 $\mu$m on average. 

Using Eq. (1), $n^2$ is 289$^2$ for 1 $\mu$m radius defect  when the distance  between the volume defects in pinned state ($L$) 
 is 3.9 $\mu$m (which is comparable with $L$= 50 $\mu$m of pure MgB$_2$) and H$_{c1}'$ is 2000 Oe. 289$^2$ exceeds flux-pinning limit of  1 $\mu$m radius defect, which is 276$^2$, thus  1 $\mu$m radius defect is able to pin fluxes up to  276$^2$ ones. 
On the other hand, 
 we can calculate the number of pinned fluxes at 1 $\mu$m radius defect from  the diamagnetic property of the specimen at 2000 Oe by using Eq. (5), which is 5 wt.\% (Fe, Ti) particle-doped MgB$_2$ as shown in Fig. \ref{fig0} (c). It was considered that $m$ of Eq. (4) is 1 because  1 $\mu$m radius defect  can pin up to 276$^2$ fluxes whereas 0.163 $\mu$m radius defect can pin up to 45$^2$ ones, which are scattered around the volume defect of 1 $\mu$m radius. As the result of calculation, 
fluxes pinned at the defect of 1 $\mu$m radius were pick-out depinned when 230$^2$ fluxes are pinned at the defect, which is contrast with 130$^2$ ones of pure MgB$_2$. Therefore,  $\Delta$G$_{dynamic}$ of pinned fluxes at 1$\mu$m radius volume defect of  5 wt.\% (Fe, Ti) particle-doped MgB$_2$ specimen is much smaller than that of the pure MgB$_2$. 

On the other hand, calculating $n^2$ of  163 nm radius defect by Eq.(1) under the same conditions (L=3.9 $\mu$m and H$_{c1}'$=2000 Oe), we obtained that $n^2$ is 68$^2$. However, 163 nm radius defect has the flux-pinning limit which is 45$^2$ when H$_{c2}$ of the superconductor is assumed to be 50 T. 
Thus, the difference between the number of pinned fluxes calculated from $\Delta$G$_{defect}$ and the flux-pinning limit by H$_{c2}$ is considerable at H$_{c1}'$ ($\Delta$G$_{pinning}$$\gg$$\Delta$G$_{pickout}$), which 
means it is little affected by $\Delta$G$_{dynamic}$. $\Delta$G$_{dynamic}$ itself is small because 163 nm radius defect can pin small number of quantum fluxes (45$^2$). Therefore, the number of fluxes pinned at 163 nm radius defect is hardly affected by $\Delta$G$_{dynamic}$ when they are pick-out depinned for the defect. 

\section{Conclusion}

We experimentally showed an increase of diamagnetic property by flux-pinning effect of (Fe, Ti) particle in MgB$_2$ which is a volume defect-dominating superconductor. And we theoretically proved the phenomenon and suggested relative equations.
A new critical field called H$_{c1}'$ was suggested, which is defined as a field showing the maximum diamagnetic property in the real superconductor. 
As a results of calculation, H$_{c1}'$ depends on  radius and the number of volume defect in superconductor which cause flux-pinning effects. When volume defects are large and a few, the experimental results are much lower than the calculated ones because of a large $\Delta$G$_{dynamic}$ which is origin of flux jump. However, when  volume defects are small and many,  the experimental results are closer to the calculated ones because of a small $\Delta$G$_{dynamic}$.  

\newpage

\vspace{2cm}

 $\bf{Acknowledgment}$\\
The authors would like to thank Dr. B. J. Kim of PNU for careful discussion.\\



\begin{figure}
\vspace{1cm}
\begin{center}
\includegraphics*[width=16cm]{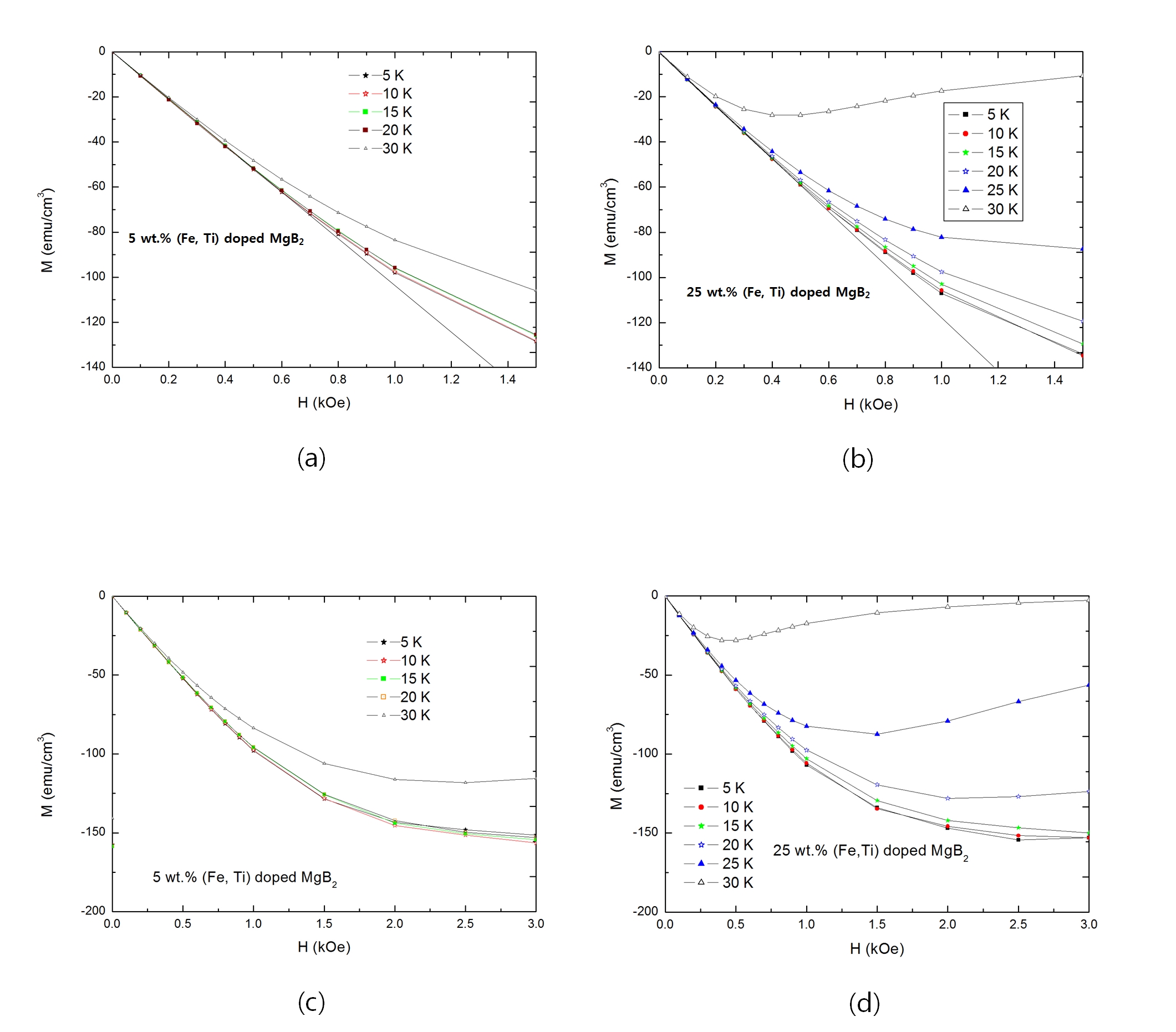}
\end{center}
\caption{Diamagnetic behaviors of volume defects-dominating superconductors around H$_{c1}$. Deviated point from the linear line is  H$_{c1}$. (a): Diamagnetic behaviors of 5 wt.$\%$ (Fe, Ti) particle-doped MgB$_2$ with various temperature. It is determined that H$_{c1}$ is  approximately 600 Oe at 0 K. (b): Diamagnetic behaviors of 25 wt.$\%$ (Fe, Ti) particle-doped MgB$_2$ with various temperature. It is determined that H$_{c1}$ is  approximately 500 Oe at 0 K, which is rather lower than that of 5 wt.$\%$ (Fe, Ti) particle-doped MgB$_2$. Increases of diamagnetic property  after H$_{c1}$  are shown in both specimens. (c): Diamagnetic behaviors of 5 wt.$\%$ (Fe, Ti) particle-doped MgB$_2$ which are extended to 3.0 kOe. (d): Diamagnetic behaviors of 25 wt.$\%$ (Fe, Ti) particle-doped MgB$_2$ which are extended to 3.0 kOe. It is observed that (c) and (d) both have a broad H$_{c1}'$ as applied magnetic field H increases, which means $\Delta$H=$\Delta$B region. }
\label{fig0}
\end{figure}

\newpage

\begin{figure}
\vspace{1cm}
\begin{center}
\includegraphics*[width=14cm]{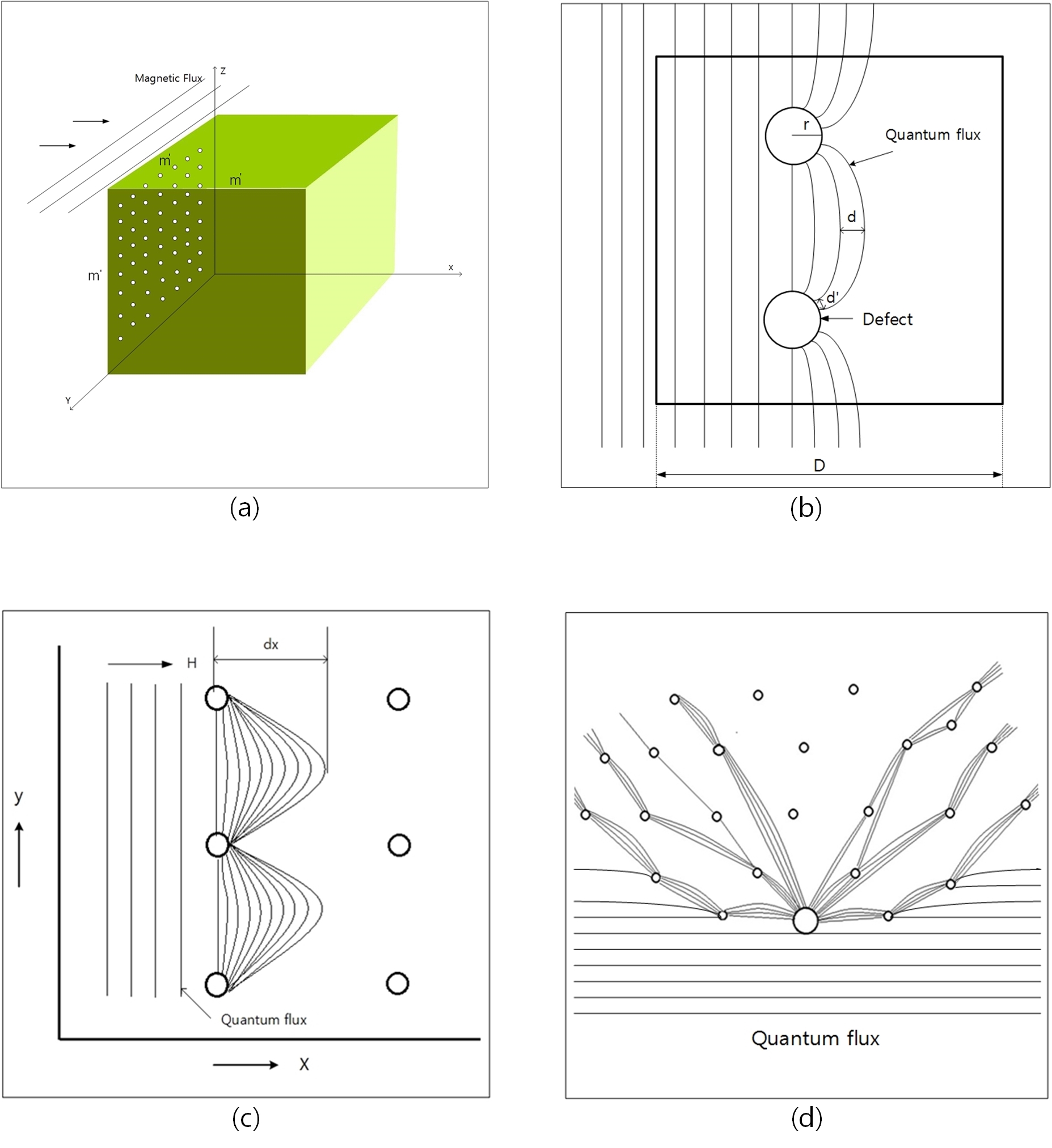}
\end{center}
\caption{ Schematic representations of flux-pinning on volume defects. (a): A distribution of volume defects in a superconductor. It is assumed that an each axis has m$'$ defects of radius r in 1 cm$^3$ superconductor. 
 If applied magnetic field exceeds H$_{c1}$, penetrated fluxes are pinned on volume defects existing from the surface of the superconductor. (b): A definition of d and d$'$. (c): A schematic representation of several quantum fluxes pinned on defects simultaneously. 
(d): A schematic representation that fluxes are pinned on several defects simultaneously when large and small volume defects are present together in a superconductor.} 
\label{fig1}
\end{figure}

\newpage
 \begin{figure}

\begin{center}
\includegraphics*[width=12cm]{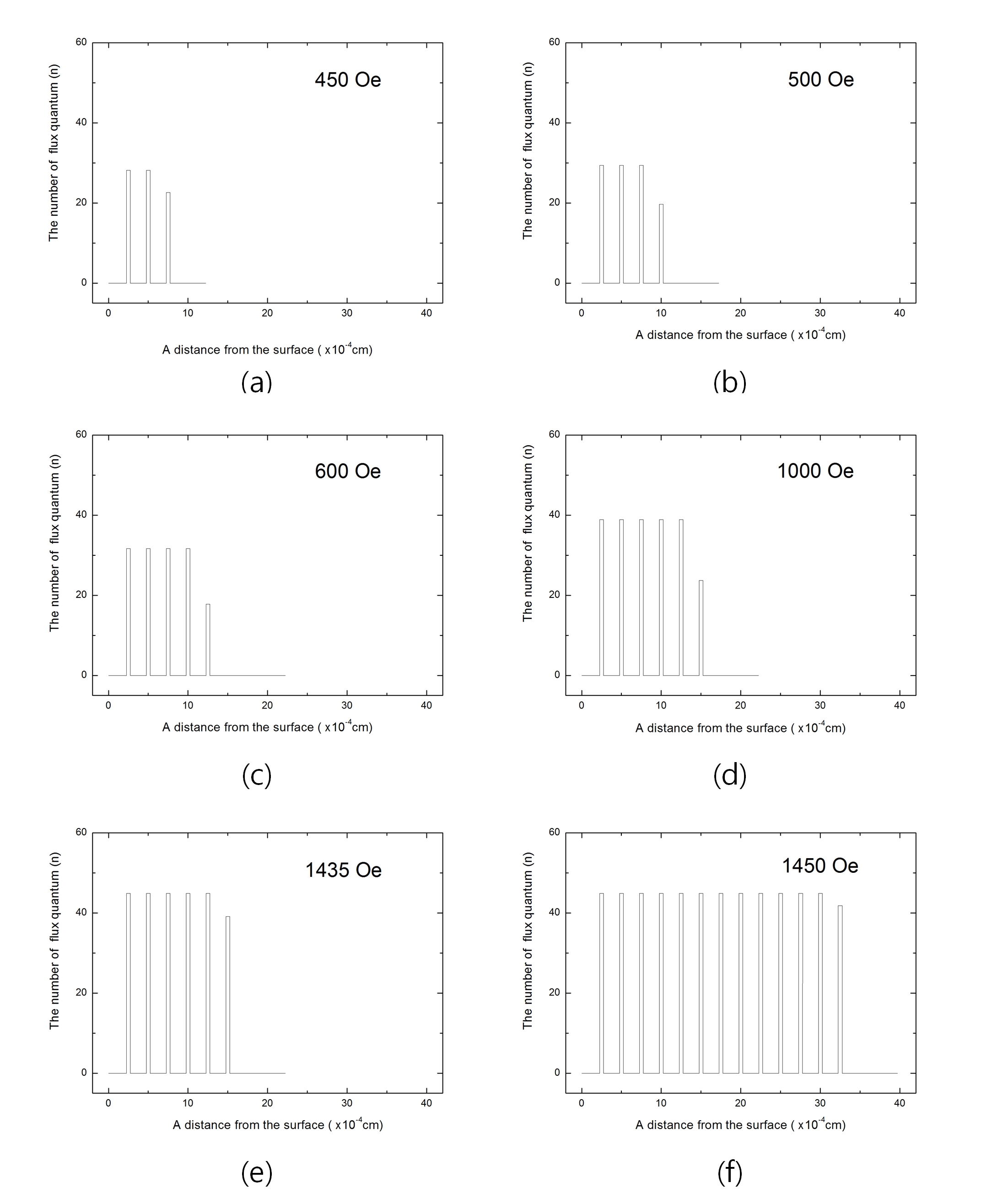}
\end{center}
\caption{ Flux-penetrating methods in volume defect-dominating superconductor 
and  the number of flux quanta pinned at a defect along applied magnetic field. It is assumed that the superconductor 4000$^3$ volume defects of which radius is 0.163 $\mu$m  in 1 cm$^3$.  Vol.\% of volume defects is 0.29, H$_{c1}$ is 400 Oe and calculated H$_{c1}'$ is 1435 Oe. (a): Calculated number of flux quanta pinned on volume defects when H is 450 Oe. Penetrated fluxes are pinned to third volume defect. (b): Calculated number of flux quanta pinned on volume defects when H is 500 Oe.  Penetrated fluxes are pinned to fourth volume defect. 
(c): Calculated ones  when H is 600 Oe. (d): Calculated ones when H is 1000 Oe. (e): Calculated ones  when H is 1435 Oe. Each volume defect reachs its flux-pinning limit. (f): Calculated ones when H is 1450 Oe. $\Delta$H=$\Delta$B region are formed because volume defects already reached their flux-pinning limit.}
\label{fig2-6}
\end{figure}

\newpage
\begin{figure}
\begin{center}
\includegraphics*[width=16cm]{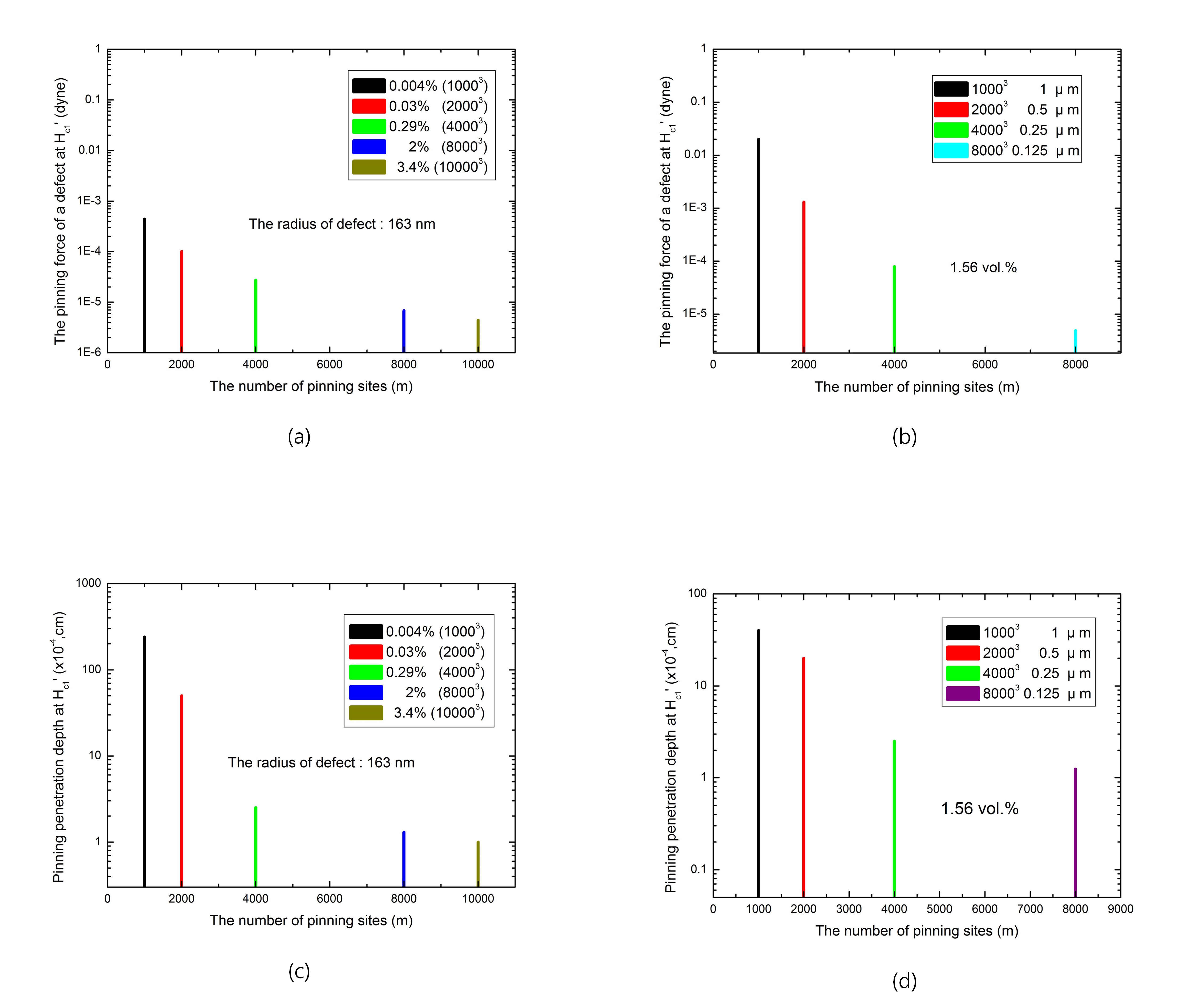}
\end{center}
\caption{Pinning forces of a volume defect at H$_{c1}'$ and pinning penetration depth at H$_{c1}'$ when H$_{c1}$ is 600 Oe,  pinning penetration depth is defined as the depth of penetrated fluxes in superconductor at H$_{c1}'$. (a): Calculated pinning forces of a volume defect at H$_{c1}'$ along  number of volume defects when a radius of a volume defect is 163 nm.  (b): Calculated pinning forces of a volume defect at H$_{c1}'$ along the number of  volume defects when vol.$\%$ of  volume defects are 1.56. (c): Calculated pinning penetration depth at H$_{c1}'$ along the number of volume defects in superconductor when radius of  volume defects is 163 nm. (d): Calculated pinning penetration depth at H$_{c1}'$ along the number of volume defects in superconductor when vol.$\%$ of volume defects are 1.56.} 
\label{fig2-7}
\end{figure}

\begin{figure}
\vspace{2cm}
\begin{center}
\includegraphics*[width=12cm]{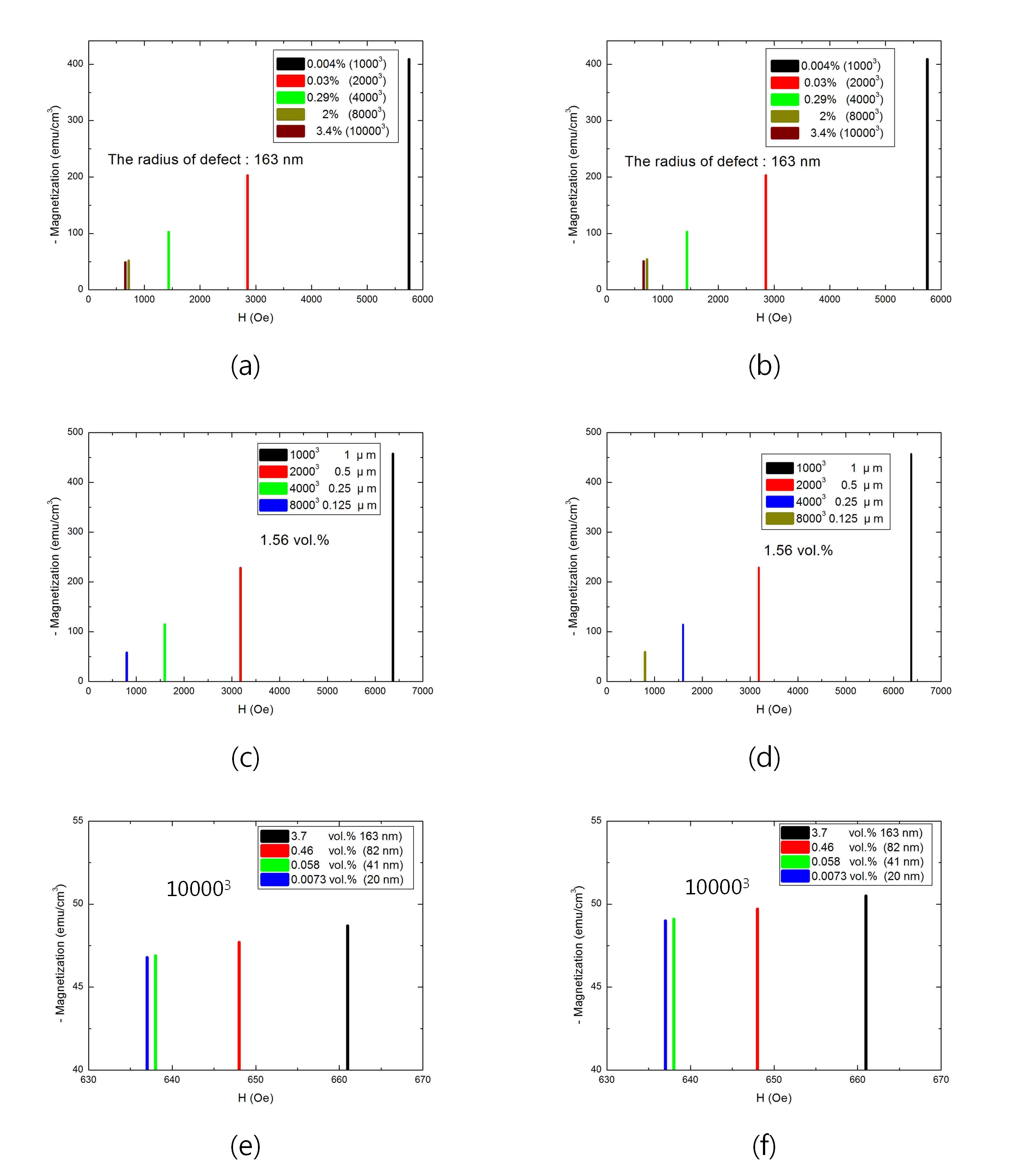}
\end{center}
\caption{ Calculated diamagnetic properties at H$_{c1}'$ for various conditions, which shows the maximum diamagnetic property in a superconductor. (a): Calculated diamagnetic properties at H$_{c1}'$ according to  vol.\% of volume defects when the radius of volume defects is 163 nm and  H$_{c1}$ is 400 Oe.  
(b): 
 Same behaviors as (a) except that H$_{c1}$ is 600 Oe.  (c): Calculated diamagnetic properties at H$_{c1}'$ when the concentration of volume defects is 1.56 vol.\% and H$_{c1}$ is 400 Oe. (d): Same behaviors as (c) except that H$_{c1}$ is 600 Oe. (e): Calculated diamagnetic properties at H$_{c1}'$ for various vol.\% of defects when the number of volume defects are fixed as 10000$^3$ and H$_{c1}$ is 400 Oe. (f): Same behaviors as (e) except that H$_{c1}$ is 600 Oe.  }
\label{fig7}
\end{figure}

\begin{figure}
\vspace{2cm}
\begin{center}
\includegraphics*[width=13cm]{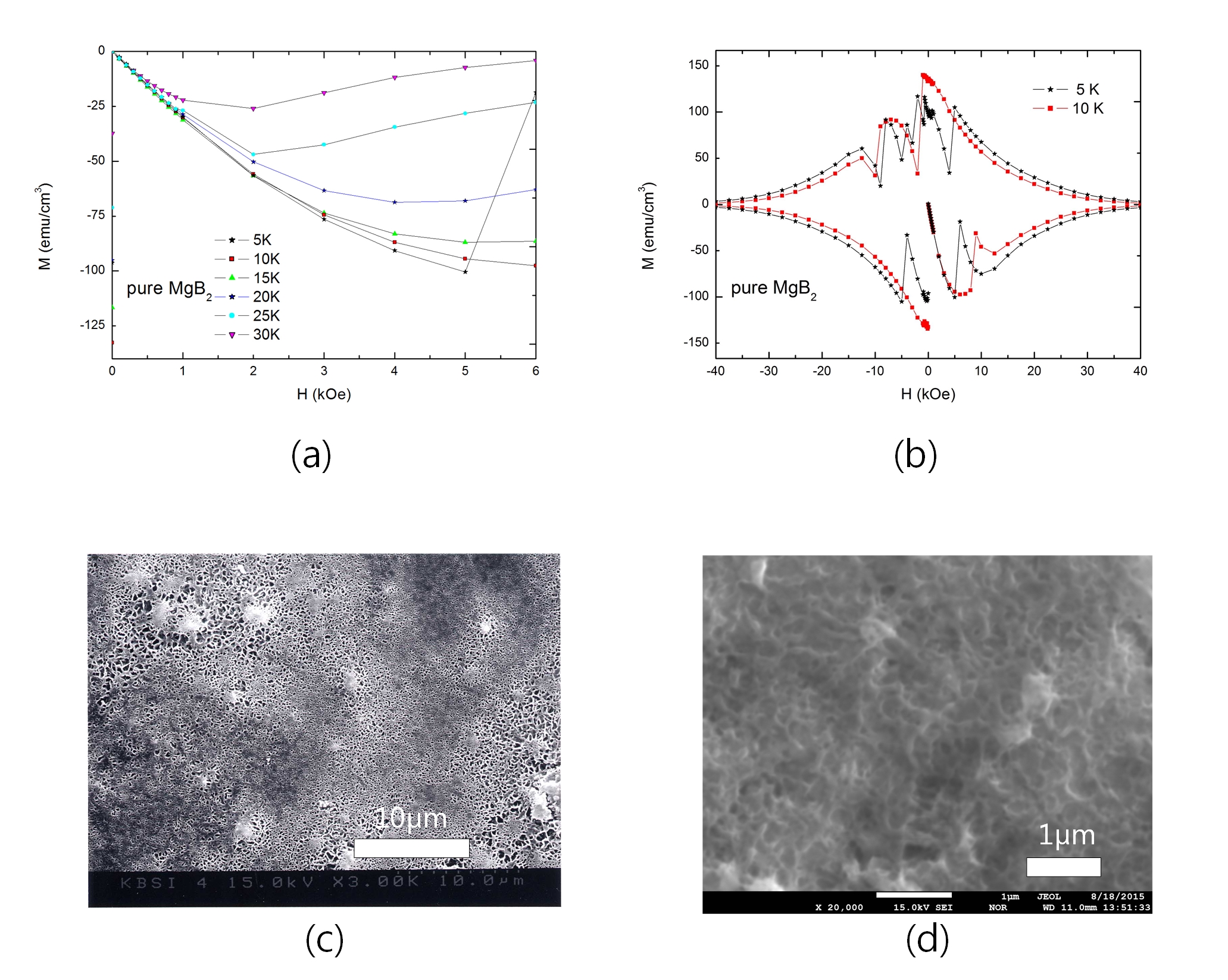}
\end{center}
\caption{ M-H curves and photographs of pure MgB$_2$ and (Fe, Ti) particle-doped  MgB$_2$. (a): M-H curves of pure  MgB$_2$ to 6 kOe. A flux jump was observed at 5 kOe on the M-H curve at 5 K. (b): Full M-H curves of pure MgB$_2$ at 5 K and 10 K. 
(c): A photograph of pure MgB$_2$. The white bright ones in MgB$_2$ base are volume defects. (d): A photograph of 5 wt.$\%$ (Fe, Ti) particle-doped  MgB$_2$. (Fe, Ti) particles are observed in MgB$_2$ base.} 
\label{fig8}
\end{figure}

\end{document}